# HOW AND WHEN TO FLATTEN JAVA CLASSES?

Jehad Al Dallal

Department of Information Science, P.O. Box 5969, Safat 13060, Kuwait

## ABSTRACT

*Improving modularity and reusability are two key objectives in object-oriented programming. These objectives are achieved by applying several key concepts, such as data encapsulation and inheritance. A class in an object-oriented system is the basic unit of design. Assessing the quality of an object-oriented class may require flattening the class and representing it as it really is, including all accessible inherited class members. Thus, class flattening helps in exploring the impact of inheritance on improving code quality. This paper explains how to flatten Java classes and discusses the relationship between class flattening and some applications of interest to software practitioners, such as refactoring and indicating external quality attributes.*

## 1. INTRODUCTION

Object-oriented programming paradigm forces the implementation of several key features, such as data encapsulation and inheritance. Properly implementing these two object-oriented concepts results in high-quality systems [1]. A class in an object-oriented system is the basic unit of design, and it encapsulates a set of attributes and methods. Data encapsulation refers to packing the strongly related attributes and methods within an object of a class and enforcing accessibility restrictions to these attributes and methods. Inheritance is a mechanism that allows for implementing the is-a relationship. Instead of redefining the attributes and methods that are included in other classes, a class can inherit these attributes and methods and only implement its unique attributes and methods, which results in reducing code redundancy and improving code testability and maintainability [2]. Representing a class with its inherited attributes and methods is called class flattening [3].

Researchers have divided class quality attributes into internal and external categories [4]. External quality attributes are attributes that indicate the quality of the class based on factors that cannot be measured using only knowledge of the software artifacts. For example, class maintainability cannot be measured unless the class is actually maintained. Similarly, class reusability cannot be measured unless the class is actually reused. It is difficult to anticipate the number of classes that will reuse the class in the future or to measure the effort that is required to maintain the class. On the other hand, internal class quality attributes, such as size, coupling, and cohesion, are attributes that can be measured based on only the knowledge of class artifacts. The internal quality attributes are not of interest for software practitioners unless they are used to indicate external quality attributes or perform activities of interest, such as refactoring [4].

Researchers have identified several external quality attributes, such as adaptability, reusability, understandability, maintainability, testability, and completeness, as somehow related to internal quality attributes [5]. Adaptability refers to the extent to which software or part of it adapts to change in environments other than those for which it was specifically designed. Reusability refers to the ease with which a component can be used in building other components. Understandability, maintainability, and testability refer to the ease with which a given piece of software can be understood, maintained, and tested, respectively. Completeness refers to the extent to which a piece of software is complete, in terms of capabilities.

In the context of this paper, we are concerned with whether a software engineer must consider class flattening before using size, cohesion, and coupling metrics to indicate the adaptability, reusability, understandability, maintainability, testability, and completeness of a subclass. We explain how to perform class flattening for Java classes. Our explanation considers the related key concepts of object-oriented programming, such as data encapsulation, overriding, and overloading. In addition, we discuss whether it is better to apply the class flattening process when using the metrics that measure the internal quality attributes in applications of interest for software practitioners, including refactoring and indicating external quality attributes. As a result, the paper answers the following two research questions:

RQ1: How to flatten java classes?
RQ2: When to flatten Java classes?

This paper is organized as follows. Section 2 reviews related work. Section 3 explains how to flatten Java classes. Section 4 discusses the applications of class flattening. Finally, Section 5 concludes the paper and discusses future work.

## 2. RELATED WORK

Bieman and Kang [6] proposed three options for inherited methods and attributes when performing cohesion analysis. The options are (1) including both, (2) excluding both, and (3) including attributes and excluding methods. Briand et al. [7] added the fourth option, which is excluding attributes and including methods. Some authors [6, 7, 8, 9] theoretically addressed the influence of inheritance on their proposed metrics and discussed whether to consider inheritance in the calculation of cohesion. However, they left the empirical investigation of this issue open for further research. Beyer et al. [3] studied the impact of inheritance on class size, cohesion, and coupling. However, their study has several limitations. First, they used relatively small numbers of classes in their empirical study, which reduces the confidence in their results. Second, they examined only one option of inheritance, which is including both inherited attributes and methods. Third, they considered the impact of inheritance on only five quality metrics. Fourth, they did not discuss the impact of class flattening on the applications of interest for software practitioners, such as refactoring and indicating external quality attributes. Chhikara et al. [10] represented a more limited analysis for the impact of inheritance on quality attributes. They considered four different designs of an artificial class example, measured the quality of each design using the six metrics proposed by Chidamber and Kemerer [11], and compared the obtained quality values. They found that the design with the best inheritance architecture has the best quality values. Several researchers empirically explored the impact of internal quality attributes on external ones [12, 13, 14, 15, 16, 17, 18, 19, 20]. This paper extends our conference-based paper [21] by providing more details regarding the methodology required to perform the class flattening and elaborating more on the class flattening applications.

## 3. HOW TO FLATTEN JAVA CLASS

This section is related to RQ1 and it explains how to consider Java semantics when performing class flattening process. Inheritance is a key concept in object-oriented programming. Implementing this concept implies distributing the features among the classes that have inheritance relations. Generally, a subclass inherits the members of its direct and transitive superclasses. The members of a class are its attributes and methods. A subclass c1 transitively inherits a superclass c2 when the subclass c1 directly inherits a superclass c3 that directly or transitively inherits the superclass c2. Syntactically, Java reserves the keyword extends to indicate the inheritance relation and allows for single inheritance only. By default, a class without a declared superclass inherits the Object class. Flattening a class refers to the process of representing the class as it really is, which means considering all of its inherited attributes and methods [3]. Java has some semantics that must be accounted for when performing class flattening. These semantics are related to data encapsulation, overriding, overloading, and dynamic binding concepts.

### 3.1. Inheritance Related Concepts in Java

Java provides four accessibility levels for class members: public, package, protected, and private. A subclass can directly access its superclass members for any accessibility level except for private. Private members in a class can only be directly accessed within the class in which they are declared. Typically, developers are advised to declare the attributes to be private and make them accessible only through access methods (i.e., setters and getters). In this paper, we refer to the private attributes and methods as invisible attributes and methods. The rest of the attributes and methods are visible.

In Java, attribute overriding occurs when an attribute of a subclass has the same name as an attribute in the direct or transitive superclass, regardless of the types of the two attributes. In this case, the subclass can access the superclass attribute either indirectly through the access methods of the attribute or directly by using the keyword super (given that the superclass attribute is not declared private). Method overriding occurs when a method in the subclass and a method in the superclass have identical signatures (i.e., method name and types, number, and ordering of the parameters). Similar to the overridden attributes, overridden methods can be accessed indirectly by calling a method (in the superclass) that invokes the overridden method or by using the keyword super. It is important to note that Java does not allow overriding an attribute when the attribute is declared static in either the subclass or the superclass. That is, both of the attributes must be similarly declared as either static or nonstatic. The same constraint applies for method overriding. Final attributes and methods cannot be overridden.

Overloading refers to the existence of two or more methods with the same names but different numbers, types, or ordering of parameters. The methods can be within the same class or in different classes with inheritance relations. In other words, Java allows for a method in a subclass to overload a method in a superclass. Dynamic method binding refers to resolving the references to subclass methods at runtime. The class flattening considered in this paper does not consider dynamic method binding because our analysis is performed statically, whereas studying the impact of dynamic method binding requires analyzing the classes dynamically. In the following, we will discuss the impact of considering these Java concepts on attribute and method flattening. Attribute flattening refers to the process of pulling down the attributes of the superclass to the subclass during the class flattening process. Similarly, method flattening refers to the process of pulling down the methods of the superclass to the subclass during the class flattening process. In case a chain of superclasses exists, class flattening is applied for each pair of subclass and superclass, starting from the subclass that inherits the superclass with no declared superclass. For example, if class c1 inherits class c2, which in turn inherits class c3, then class flattening will be applied for class c2 first. The resulting flattened version of class c2 will be used in the class flattening process of class c1.

### 3.2. Attribute Flattening

Nonoverridden attributes can be either visible or invisible. In the attribute flattening process, the nonoverridden attributes that are visible must be pulled down from the superclass to the subclass because these attributes are accessible through the subclass. A nonoverridden attribute that is invisible can be either accessed by some methods in the superclass or not. In the former case, the attribute must be pulled down to the subclass in the flattening process unless none of the methods that access the attribute in the superclass are pulled down to the subclass in the method flattening process. A compilation error will result if the invisible attribute is not pulled down to the subclass, and, at the same time, a method that accesses the attribute is pulled down to the subclass in the flattening process. In the flattening process, the nonoverridden attribute that is declared private will not be pulled down to the subclass if the attribute is not accessed by any of the methods in the superclass. Such an attribute is invisible and inaccessible. The existence of such an attribute is syntactically correct but meaningless.

Regardless of its visibility, an overridden attribute must be renamed and pulled down to the subclass if it is accessed by some methods that are pulled down to the subclass in the flattening process; otherwise, the flattened class will have a compilation error. The pulled down methods that access the pulled down attribute and the original subclass code that accesses the pulled down attribute using the super keyword must be modified to refer to the new name of the attribute. Otherwise, the pulled down methods or the subclass code will be incorrectly accessing the overriding attributes instead of the overridden ones. On the other hand, if the overridden attribute is visible and it is not accessed by any of the methods in the superclass, then the attribute must be renamed and pulled down to the subclass. In this case, the attribute is pulled down because it was allowed to use the keyword super to access the attribute in the original version of the subclass. The attribute is renamed to avoid a naming conflict, which causes a compilation error. The subclass code that accesses the pulled down attribute must be modified to refer to the new name of the attribute. Finally, an overridden attribute that is invisible will not be pulled down to the subclass if the attribute is not accessed by any of the methods that are pulled down during the flattening process. As discussed above, this attribute is invisible and inaccessible, and its existence in the superclass indicates an anomaly case.

In summary, unless the attribute is invisible and not accessed by any of the superclass methods that are pulled down during the flattening process, the attribute must be pulled down to the subclass during the flattening process. The pulled down overridden attributes must be renamed, and the corresponding code must be modified.

### 3.3. Method Flattening

The above discussion regarding attribute flattening is also applicable for the superclass methods considered in the method flattening process. That is, invisible methods that are not accessed by any of the visible methods of the superclass must not be pulled down to the subclass during the class flattening process. These methods are meaningless. All of the other visible and invisible methods must be pulled down to the subclass for the same reasons mentioned above regarding the attributes. Pulled down overridden methods must be renamed, and the corresponding code must be modified. The diagram given in Figure 1 summarizes the different possibilities, discussed above, and the corresponding flattening actions.

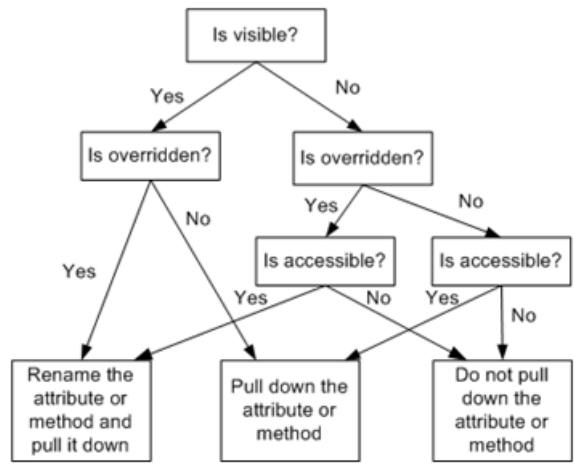

Figure 1. The different possibilities of the attribute or method cases and the corresponding flattening actions

## 4. WHEN TO APPLY CLASS FLATTENING

This section is related to RQ2 and it discusses whether class flattening must be performed before assessing class quality. The considered quality attributes include adaptability, reusability, understandability, maintainability, completeness, and testability. In addition, we discuss whether class flattening must be performed before applying several refactoring activities.

A subclass cannot be moved alone to a new environment. Instead, all of the direct and indirect superclasses must be also moved with the subclass to the new environment. Therefore, when studying the adaptability of a subclass, the code of the direct and indirect superclasses must be considered as well. Similarly, the reusability of the direct and indirect superclasses must be considered when reusing a subclass. A subclass cannot be understood and maintained unless the direct and indirect superclasses are inspected and analyzed. A subclass is not complete by itself without the inherited code. Therefore, studying the completeness of a subclass must include exploring the completeness of the code of the subclass and its superclasses altogether. As a result, none of these applications requires solely analyzing or exploring the code within the subclass; they also require the analysis and exploration of the direct and indirect superclasses. Therefore, it is expected that if the internal quality metrics are applied on the subclasses without considering the inherited attributes and methods, then the obtained quality values will provide incorrect indications for the external quality attributes. However, empirical studies are required to approve or disapprove this expectation.

Object-oriented testing is performed at several levels. At the class level, the code within the class is tested. At the cluster level, the relations between related classes are tested. Therefore, we expect that when applying the internal quality metrics to the subclasses without being flattened, the values obtained can be somehow used to indicate the class testability. To indicate the testability of classes with inheritance relations (i.e., testing at the cluster level), software engineers must consider the flattened versions of the subclasses.

Refactoring is an application of interest for software practitioners. It refers to the process of restructuring software source code to enhance its quality without affecting its external behavior [22]. Fowler [22] identified several refactoring scenarios such as Extract Class, Extract Subclass, Extract Superclass, and the Move Method. In the Extract Class refactoring activity, a class with the association relation to the original class is extracted. In the Extract Subclass refactoring activity, a subclass is created to include a subset of the features of the original class. In the Extract Superclass refactoring activity, a superclass is created to include the common features of several classes. In the Move Method refactoring activity, a method is moved to the class that uses the method the most. Several techniques are proposed in the literature to identify the refactoring opportunities and to perform the refactoring activities. Some of these techniques are based on measuring the code's internal quality attributes [23, 9]. It is expected that these techniques will not function properly if the flattened version of the classes are considered. For example, assume that a software engineer would like to test whether a class c1 is in need of Extract Subclass refactoring and that this class of interest is a subclass of another class c2. The flattened version of class c1 will almost completely include the code of both classes c1 and c2 (with the constraints discussed in Section 3). Therefore, if the software engineer applies the technique that indicates whether class c1 is in need of refactoring on the flattened version of c1, then the technique will incorrectly indicate that the class is in need of Extract Subclass refactoring. As a result, when using refactoring techniques that are based on assessing the internal quality attributes, the software engineer must pay attention to the impact of class flattening on the intended refactoring activity. Otherwise, incorrect refactoring decisions will be taken.

## 5. CONCLUSIONS AND FUTURE WORK

This paper demonstrates how to flatten Java classes and explains which of the superclass attributes and methods must be considered in class flattening. It shows how to perform attribute and method flattening to obtain different views of the class that inherits other classes. We argued that it is better to measure the internal quality attributes using the flattened versions of the classes when using the resulting quality values to indicate the class adaptability, reusability, understandability, and maintainability. In contrast, it is better to use the original versions of the classes when using the resulting quality values in class refactoring processes or to indicate the class testability. To prove or disapprove our theoretically based expectations, in the future, we plan to empirically study the impact of class flattening when using the internal quality attributes to indicate the external quality attributes.

### ACKNOWLEDGEMENTS

The author would like to acknowledge the support of this work by Kuwait University Research Grant QI02/13.

**Author**

Jehad Al Dallal received his PhD in Computer Science from the University of Alberta in Canada and was granted the award for best PhD researcher. He is currently working at Kuwait University in the Department of Information Science as an Associate Professor and Department Chairman. Dr. Al Dallal has completed several research projects in the areas of software testing, software metrics, and communication protocols. In addition, he has published more than 80 papers in reputable journals and conference proceedings. Dr. Al Dallal was involved in developing more than 20 software systems. He also served as a technical committee member of several international conferences and as an associate editor for several refereed journals.